\documentclass[twocolumn]{jpsj3} 
\newcommand{\nn}{\nonumber}

\def\runauthor{Online-Journal Subcommittee}

\title{%
Stability of the Singular Vortex and Associated Majorana Zero Modes 
in Trapped $p$-Wave Resonant Superfluids of Neutral Cold Atoms
}

\author{%
Yasumasa \textsc{Tsutsumi}\thanks{E-mail address: tsutsumi@mp.okayama-u.ac.jp} and Kazushige \textsc{Machida}
}

\inst{%
Department of Physics, Okayama University,
Okayama 700-8530, Japan
}

\recdate{\today}

\abst{%
The stability conditions for the singular vortex that accompanies Majorana zero modes at the core 
are investigated for $p$-wave resonant superfluids of atomic Fermi gases.
Within the Ginzburg-Landau framework, we determine the stable conditions in the parameter space 
for the external rotation frequency and the harmonic trap frequency.
There exists a narrow stable region in this parameter space for quasi-two-dimensional condensates.
We also describe the detailed characterizations of the spatial structure of the order parameter 
in the chiral $p$-wave superfluids under rotation.
}

\kword{chiral $p$-wave pairing, superfluids, atomic Fermi gases, Majorana zero modes, vortex structures
}

\begin{document}
\maketitle

\section{Introduction}
Chiral triplet superconductors and superfluids described symbolically by
$p+ip$ pairing, which breaks time reversal symmetry, are a typical 
topological order and possess many interesting and useful nontrivial physical properties if those really exist.
As for superconductors, Sr$_2$RuO$_4$ is often referred to as a prime candidate chiral triplet superconductor.
However, the precise pairing symmetry has not yet been identified and has been under strong discussion. 
A naive $p_x+ip_y$ scenario is reexamined from various aspects \cite{214}.
Therefore, at this moment, there is no concrete material 
that firmly exhibits chiral triplet pairing symmetry in a superconductor \cite{UPt3}.

As for superfluids, the $^3$He-A phase is definitely in the chiral pairing state, which
has been a testing ground for checking ideas associated with the topological order \cite{vollhardt}.
However, because of its inherent strong interacting nature from the theoretical
side and the limited accessibility from the experimental side, it is a difficult and challenging task to explore superfluid $^3$He.
Needless to say, we wish to find more clean 
systems of chiral superfluids that are easily accessible theoretically and experimentally.

In this sense, the use of neutral atomic Fermi gases is another option to pursue this possibility.
By using magnetic Feshbach resonance, a $p$-wave interaction channel can be enhanced,
leading to $p$-wave paired superfluids. 
In fact, there have been vigorous activities toward realizing $p$-wave
resonant superfluids in $^6$Li \cite{inada} and $^{40}$K \cite{jin} recently.
The advantages of $p$-wave resonant superfluids obtained using atomic Fermi gases are
(1) the tunability of the attractive interaction strength in a $p$-wave channel from strong to weak cases,
(2) the controllability of the system dimension by changing a trap potential
from three dimensions (3D) to two dimensions (2D), even to one dimension (1D),
and (3) ease of manipulations, such as rotation, and excitations of collective modes by external perturbations.

Chiral $p$-wave superfluids have nontrivial Caroli-de Gennes-Matricon (CdGM) states \cite{CdGM} at a vortex core.
It is possible for the CdGM states to have exact zero energy states in the BCS regime \cite{volovik,tewari1,mizushima}.
Since the zero energy states can be described by the 1D Majorana equation \cite{tewari2},
the zero energy states are Majorana zero modes.
A remarkable feature of Majorana zero modes is that their creation operator is expressed by the self-Hermitian operator
$\gamma=\gamma^{\dagger}$ and their host vortices obey the non-Abelian statistics \cite{ivanov,read}.
These nontrivial properties can be utilized for a topological quantum computer \cite{kitaev}.

A pair of conventional fermionic creation and annihilation operators span a 2D Hilbert space because their square vanishes.
This is not true for the Majorana operators.
Thus, to avoid the problem when $2N_0$ vortices are present,
we can construct ``conventional" complex fermionic creation and annihilation operators,
$\psi_i^{\dagger}=(\gamma_i-i\gamma_{N_0+i})/2$ and $\psi_i=(\gamma_i+i\gamma_{N_0+i})/2$, respectively,
from a pair of vortices,
where the normalization is chosen as $\gamma_i^2=1$ and subscript $i$ denotes the index of vortices \cite{nayak}.
These operators satisfy $\psi_i^2=\psi_i^{\dagger 2}=0$ 
and thus span a 2D subspace of degenerate ground states associated with these operators.
Strictly speaking, a topological vortex qubit is defined through two pairs of vortices
because the fermionic number is fixed to be even or odd \cite{zhang}.

A necessary condition for Majorana zero modes to exist at a vortex core is as follows: 
(A) the vortex possesses an odd winding number and phase singularity in spinless chiral $p$-wave superfluids 
\cite{tsutsumi1,mizushima} or
(B) the vortex possesses one-half winding number, that is, the vortex is a half-quantum vortex
in spinful chiral $p$-wave superfluids \cite{ivanov}.

Here, under those circumstances, 
we investigate experimental requirements to satisfy the condition (A) in neutral atomic Fermi gases.
Since the population of atoms in each hyperfine spin state is controllable by an RF field, 
the spinless $p$-wave superfluids could be achieved.
In quasi-2D confinement that prohibits the Cooper pairs of the $z$-direction,
$p_x \pm ip_y$ paired states are realized 
unless the magnetic field for the Feshbach resonance is parallel to the quasi-2D system \cite{gurarie,cheng}.
A requirement of the quasi-2D system is $R_z \ll \xi$, 
namely, $\hbar \omega_z/(k_BT_c) \gg T_c/T_F$,
where $R_z$ is the radius of atomic gases toward the $z$-direction, $\xi$ is the coherence length, 
$\omega_z$ is the trap frequency of the $z$-direction, $T_c$ is the transition temperature, and $T_F$ is the Fermi temperature.
This requirement may be satisfied by the optical lattice potential of a period $\sim 3$ $\mu$m \cite{dalibard}.
The aim of this study is to determine the stability condition for the singular vortex,
which possesses Majorana zero energy modes at the vortex core, under the quasi-2D confinement 
that allows the chiral $p_x\pm ip_y$ paired state produced by the magnetic Feshbach resonance.
It has been proposed that the vortices with Majorana zero modes in the $p$-wave resonant atomic Fermi gases 
are utilized for a topological quantum computer already 
including the optical methods of operating and reading out states of qubits \cite{tewari1,zhang}.
However, the strict stable conditions of the vortices with Majorana zero modes have not yet been discussed.

In \textsection{2}, we derive the Ginzburg-Landau (GL) functional form
within the weak coupling approximation in order to describe the $p_x \pm ip_y$ paired state under rotation,
and classify the possible vortex types in a 2D trapped system.
We construct the phase diagram for the stable singular vortex, which allows Majorana zero modes in the core,
in the parameter space, that is, rotation frequency $\Omega$ vs trap frequency $\omega$ in \textsection{3}.
The final section is devoted to conclusions.
The closely related vortex stability calculations for 3D systems  are found in refs. 20 and 21.

\section{Formulation}

\subsection{Ginzburg-Landau functional}

The order parameter (OP) of the spinless $p$-wave superfluidity in the quasi-2D trap potential 
can be expanded on the basis of rectangular coordinates with expansion coefficients $A_i$ $(i=x,y)$:
\begin{align}
\Delta(\mbox{\boldmath$p$})=A_x\hat{p}_x+A_y\hat{p}_y,
\end{align}
where $\hat{\mbox{\boldmath$p$}}$ is the unit vector in momentum space.
We assume that the momentum toward the $z$-direction is prohibited by the quasi-2D trap potential.

Here, we employ the GL framework for the $p$-wave superfluidity \cite{vollhardt}.
This framework is valid in the vicinity of the transition temperature, $T_c-T \ll T_c$,
and applicable to cold atomic Fermi gases in a harmonic trap potential with $\hbar \omega \ll k_B T_c$ \cite{baranov},
where $T_c$ and $\omega$ are the transition temperature and  the trap frequency, respectively.
The GL free energy is obtained by integrating the GL free energy density in a 2D system:
\begin{align}
F=\int d^2 r (f_{\rm bulk} + f_{\rm grad} + f_{\rm cent} + f_{\rm harm}).
\label{GL energy}
\end{align}
In the weak coupling limit,
the condensation energy, gradient energy, centrifugal energy, and trap potential energy are respectively given by
\begin{align}
f_{\rm bulk}=&- \left( 1- t \right) A_i^*A_i + {1 \over 2} A_i^* A_i A_j^* A_j + {1 \over 4} A_i^* A_i^* A_j A_j, 
\label{fbulk}\\
f_{\rm grad}=& {3 \over 5}[ (\partial_i A_j )^*(\partial_i A_j )+(\partial_i A_j )^*(\partial_j A_i )\nonumber\\
&+(\partial_i A_i )^*(\partial_j A_j ) ] , 
\label{fgrad}\\
f_{\rm cent}=&-{3 \over 5}\Omega^2r^2(A_i^* A_i + 2 |A_{\theta}|^2 ), 
\label{centrifugal}\\
f_{\rm harm}=&\quad  {3 \over 5}\omega^2r^2A_i^* A_i ,
\label{fharm}
\end{align}
where $t=T/T_c$, 
$\mbox{\boldmath$\partial$}=\mbox{\boldmath$\nabla$}-i\mbox{\boldmath$\Omega$}\times\mbox{\boldmath$r$}$
with the rotation frequency $\Omega$ around the $z$-axis,
$\mbox{\boldmath$\Omega$}\parallel\hat{\mbox{\boldmath$z$}}$,
and $A_{\theta}=-A_x\sin \theta +A_y \cos \theta$ in cylindrical coordinates.
The repeated indices $i$ and $j$ imply summations over $x$ and $y$.

The GL free energy density functionals, eqs. \eqref{fbulk} - \eqref{fharm}, are expressed with dimensionless units.
The units of OP and length are
the OP amplitude of the chiral state in a bulk, $\Delta_0=\sqrt{10\pi^2/7\zeta(3)}k_BT_c$,
and the GL coherence length $\xi_0=\sqrt{7\zeta(3)/48\pi^2}(\hbar v_F/k_BT_c)$ with zero temperature, respectively,
where $v_F$ is the Fermi velocity.
The effect of harmonic trap potential on condensates depends on the dimensionless pairing interaction $\Gamma$
that determines the transition temperature $T_c=CE_F\exp(-1/\Gamma)$,
where $E_F$ is the Fermi energy and $C$ is a numerical coefficient of order unity \cite{baranov}.
We use the  unit $\hbar\omega_0=\sqrt{144\pi^2/35\zeta(3)}\sqrt{2\Gamma / (1+2\Gamma)}k_BT_c$ 
for the trap frequency $\hbar \omega$.
The $p$-wave triplet paring interaction is $\Gamma \approx {\lambda^2 / 13}$ 
with a gaseous parameter $\lambda={2|a|p_F / \pi \hbar}$,
where $a$ is the scattering length and $p_F$ is the Fermi momentum \cite{baranov2}.
Rotation frequency is normalized by the critical rotation frequency $\Omega_c$,
where condensates fly away from the trap potential by centrifugal force.
We sum up the centrifugal energy and the trap potential energy as
$$f_{\rm cent} + f_{\rm harm}
={3 \over 5}(\omega^2-\Omega^2)r^2|A_r|^2+{3 \over 5}(\omega^2-3\Omega^2)r^2|A_{\theta}|^2$$
with cylindrical coordinates.
Therefore, the critical rotation frequency is given by $\Omega_c=\omega / \sqrt{3}$ \cite{tsutsumi2,tsutsumi3}.

We estimate the approximate profile of the condensates at rest within the Thomas-Fermi (TF) approximation.
In addition, we assume that the condensate is in the chiral state, which is stable in a bulk.
The profile of OP amplitude is
\begin{align}
|\Delta_c(r)|=\sqrt{(1-t)-{3 \over 5}\omega^2r^2},
\label{TF profile}
\end{align} 
and the radius of condensates is
\begin{align}
R_{\Delta}=\sqrt{5 \over 3}{\sqrt{1-t} \over \omega}.
\end{align} 

Here, we introduce the $l$-vector pointing in the direction of the orbital angular momentum of the Cooper pair.
The $l$-vector has only the $z$-component when the momentum toward the $z$-direction is suppressed.
The $z$-component of the $l$-vector is expressed 
with expansion coefficients on the basis of the component of the orbital angular momentum along the $z$-axis:
$\Delta (\mbox{\boldmath$p$})=A_+\hat{p}_++A_-\hat{p}_-$,
where the bases and coefficients of the plus-minus chiral states are 
$\hat{p}_{\pm}=\mp \left( \hat{p}_x \pm i\hat{p}_y \right) /\sqrt{2}$
and $A_{\pm}=\mp \left( A_x \mp iA_y \right) /\sqrt{2}$, respectively.
By using the coefficients,
\begin{align}
l_z={|A_+|^2-|A_-|^2 \over |\Delta|^2},
\end{align}
where $|\Delta|^2=|A_+|^2+|A_-|^2$ is the squared amplitude of the OP.
If the condensates are purely in the chiral state, $l_z$ has the unit value.
$l_z$ decreases when the polar state is mixed through gradient and centrifugal energies.
Finally, $l_z$ becomes zero, provided that the condensates are in the pure polar state.

We have identified stationary solutions by numerically solving the variational equation
$\delta f(\mbox{\boldmath$r$})/\delta A_i(\mbox{\boldmath$r$})=0$, 
where $f(\mbox{\boldmath$r$})$ is the GL free energy density functional, namely, the integrand of eq. \eqref{GL energy}.
In this study, we have considered the temperature $t=0.9$.

\subsection{Possible types of axisymmetric structures and vortices at low rotation}

Since we assume the axisymmetric trap potential,
axisymmetric structures of the condensates are stable 
when there is no vortex and only a singular vortex exists in the system.
Here, we define the winding numbers of the plus and minus components as $w_+$ and $w_-$, respectively,
where $A_{\pm}=|A_{\pm}|e^{i (w_{\pm}\theta + \alpha_{\pm})}$ with a phase constant $\alpha_{\pm}$.
Axisymmetric structures of the OP consisting of the plus-minus components
must have the combinations of the winding number $\langle w_+, w_- \rangle = \langle n-1,n+1 \rangle$,
where $n$ is an integer \cite{sauls1,sauls2}.
The axisymmetric condition of what is stricter than that of multicomponent spinor Bose-Einstein condensates (BECs) \cite{isoshima}.
Generally, the small winding number state is energetically favorable.

For the nonvortex structures, the possible combinations of the winding number are
$\langle w_+, w_- \rangle = \langle -2,0 \rangle$ and $\langle 0,2 \rangle$.
These structures are degenerate at rest.
However, under positive rotation or counterclockwise rotation, the structure with $\langle -2,0 \rangle$ is stable
because  the minus chiral component without phase winding is advantageous under positive rotation \cite{ichioka}.
For the structures with a singular vortex, the possible combinations of the winding number are 
$\langle w_+, w_- \rangle = \langle -3,-1 \rangle$, $\langle -1,1 \rangle$, and $\langle 1,3 \rangle$.
Among these combinations, the structure with $\langle -1,1 \rangle$ is the most stable under low positive rotation,
because the main component, which is the minus chiral component, has a winding number of 1.

Therefore, it is sufficient to consider the combination of winding number 
$\langle w_+, w_- \rangle = \langle -2,0 \rangle$ for the nonvortex structure, 
and $\langle w_+, w_- \rangle = \langle -1,1 \rangle$ for the structure with a singular vortex.

\section{Stability Region of the Singular Vortex}

\begin{figure}
\begin{center}
\includegraphics[width=8cm]{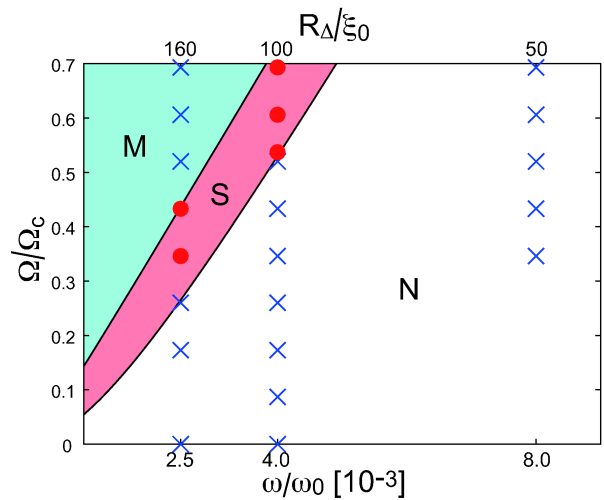}
\end{center}
\caption{(Color online) 
Phase diagram of the OP structures for trap frequency $(\omega)$ vs rotation frequency $(\Omega)$.
The TF radii of the condensate $(R_{\Delta})$ at rest are also noted.
The region N is the nonvortex structure, S is the singular vortex structure where Majorana zero modes exist, 
and M is the multiple vortex structure.
Solid circles and crosses show the structures with and without Majorana zero modes,
respectively.
} 
\label{phase}
\end{figure}

We show the phase diagram of the OP structures 
for trap frequency $\omega$ vs rotation frequency $\Omega$ in Fig. \ref{phase}.
We consider the three cases for the trap frequencies $\hbar \omega =$ $2.5$, $4.0$, and 
$8.0 \times 10^{-3}\sqrt{144\pi^2/35\zeta(3)} \sqrt{2\Gamma/(1+2\Gamma)}k_BT_c$.
The condition of applying the GL framework $\hbar \omega \ll k_BT_c$ is satisfied 
because of ${2\Gamma/(1+2\Gamma)} < 1$.
The calculated points are denoted by solid circles and crosses in Fig. \ref{phase}.
The region N is the nonvortex structure, S is the singular vortex structure where Majorana zero modes exist, 
and M is the multiple vortex region.
Details of each structure are explained in the following subsections.

\subsection{Rotation dependence}

In this subsection, we show the representative structures with the trap frequency $2.5 \times 10^{-3}\omega_0$,
where the TF radius of the condensate $R_{\Delta}$ is about $160\xi_0$ at rest.
In this case, various vortex structures are stabilized by rotation.

\begin{figure}
\begin{center}
\includegraphics[width=8cm]{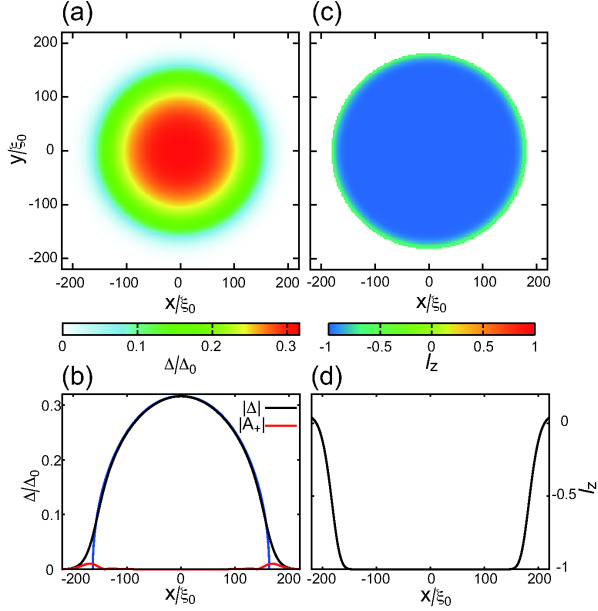}
\end{center}
\caption{(Color online) 
In the region N of Fig. \ref{phase} at rest:
(a) Contour map of the OP amplitude,
(b) cross sections of the OP amplitude, the plus chiral component, and TF estimate from eq. \eqref{TF profile},
(c) contour map of $l_z$ where the region $\Delta/\Delta_0 \ge 0.01$ is shown,
and (d) cross section of $l_z$.}
\label{N}
\end{figure}

We show the OP amplitude at rest in Figs. \ref{N}(a) and \ref{N}(b).
The OP amplitude at the center is slightly suppressed 
from that of the chiral state in a bulk $\Delta(t)=\Delta_0\sqrt{1-t}$.
The OP amplitude is gradually decreased toward the outside by the trap potential,
according to eq. \eqref{TF profile} except near the edge.
Since the condensation energy is small at the edge, the gradient energy makes a large contribution to the structures.
The plus chiral component $A_+$ with the winding number $w_+=-2$ is induced at the edge
because the gradient energy terms in eq. \eqref{fgrad}, such as $A_+^*(\partial^2A_-/\partial r^2)$, 
inevitably induce another component at the place
where the main component $A_-$ with the winding number $w_-=0$ is spatially varied.
The feature is absent in spinor BECs 
because the spatial differential operators act on real space, while the OP exists in spin space \cite{ohmi,ho}.

This is also clear from the $l_z$ values shown in Figs. \ref{N}(c) and \ref{N}(d).
In almost all regions, $l_z=-1$, namely, the condensate is in the minus chiral state.
However, the $l$-vectors shorten toward the outside of the condensate at the edge.
The condensate is finally in the polar state.

\begin{figure}
\begin{center}
\includegraphics[width=8cm]{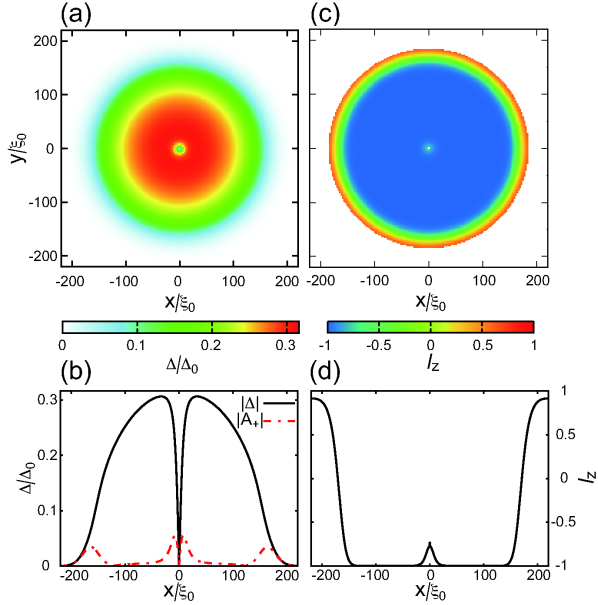}
\end{center}
\caption{(Color online) 
In the region S of Fig. \ref{phase} where a singular vortex is accompanied by Majorana zero modes 
at $\Omega=0.35\Omega_c$:
(a) Contour map of the OP amplitude,
(b) cross sections of the OP amplitude (solid line) and the plus chiral component (dot-dash line), 
(c) contour map of $l_z$ where the region $\Delta/\Delta_0 \ge 0.01$ is shown,
and (d) cross section of $l_z$.}
\label{S}
\end{figure}

A singular vortex enters the center of the condensate under rotation.
We show the OP amplitude and the $z$-component of the $l$-vector in Fig. \ref{S} at $\Omega=0.35\Omega_c$.
Since the vortex enters the condensate of the spinless chiral state and has a combination of odd winding number 
$\langle w_+, w_-\rangle = \langle -1,1\rangle$,
the vortex is accompanied by Majorana zero modes.
The condensate is in the minus chiral state in most of the regions;
however, the plus chiral component is induced around the vortex core and at the edge.
Note that the condensate at the edge is in the plus chiral state
because of the effect of rotation through the gradient energy terms in eq. \eqref{fgrad}, 
such as $\Omega rA_+^*(\partial A_-/\partial r)$.

\begin{figure}
\begin{center}
\includegraphics[width=8cm]{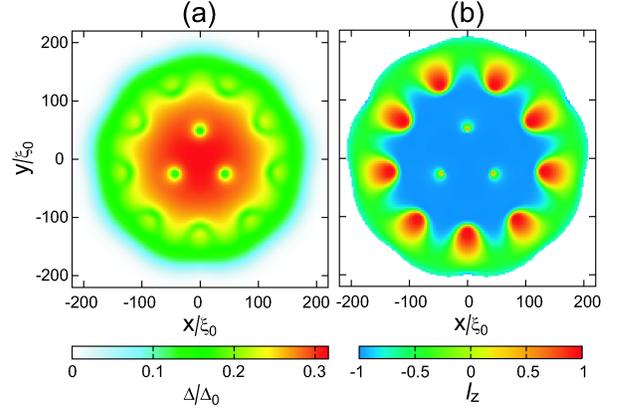}
\end{center}
\caption{(Color online) 
In the region M of Fig. \ref{phase} at $\Omega=0.60\Omega_c$:
(a) Contour map of the OP amplitude
and (b) contour map of $l_z$ where the region $\Delta/\Delta_0 \ge 0.01$ is shown.}
\label{M}
\end{figure}

On increasing the rotation frequency $\Omega$, vortices enter the condensate one by one from outside.
We show the OP amplitude in Fig. \ref{M}(a) and $l_z$ in Fig. \ref{M}(b)
where the condensate contains three vortices at $\Omega=0.60\Omega_c$.
All the vortices are not singular vortices in the system.
The cores of the plus and minus chiral components are split and covered by the other component.
The centrifugal energy, eq. \eqref{centrifugal}, increases the condensate extension
and brings the condensate at the edge into the polar state with the momentum toward the $\theta$-direction.
Half-quantum vortices with $l_z=1$ in the polar state effectively absorb the angular momentum by rotation.

Here, we introduce a pseudo-spin representation
so that differences between the singular vortex and the splitting vortex can easily be seen.
Since the OP has only two components, we can express the OP on the basis of the vector
\begin{align}
\mbox{\boldmath$\eta$}\equiv {1 \over |\Delta|}
\begin{pmatrix}
A_+ \\ A_-
\end{pmatrix}.
\end{align}
The pseudo-spin is defined using the vectors and Pauli matrices with the total spin $1/2$.
\begin{align}
S_x \equiv {1 \over 2} \mbox{\boldmath$\eta$}^{\dagger} \sigma_x \mbox{\boldmath$\eta$} = 
& {1 \over |\Delta|^2} {\rm Re} (A_+^*A_-), \nn \\
S_y \equiv {1 \over 2} \mbox{\boldmath$\eta$}^{\dagger} \sigma_y \mbox{\boldmath$\eta$} = 
& {1 \over |\Delta|^2} {\rm Im} (A_+^*A_-),  \\
S_z \equiv {1 \over 2} \mbox{\boldmath$\eta$}^{\dagger} \sigma_z \mbox{\boldmath$\eta$} = 
& {1 \over 2} {|A_+|^2-|A_-|^2 \over |\Delta|^2}. \nn
\end{align}
The $z$-component of the pseudo-spin $S_z$ is one-half of $l_z$.

We define the topological charge $N\equiv w_--w_+$.
Axisymmetric vortices have the topological charge $N=2$
and the half-quantum vortices with the combinations of the winding number 
$\langle w_+, w_- \rangle = \langle -1, 0 \rangle$ and $\langle 1, 0 \rangle$ have $N=1$.
The pseudo-spin can be rewritten with the topological charge:
\begin{align}
S_x =& {|A_+A_-| \over |\Delta|^2} \cos(N\theta +\alpha'), \nn \\
S_y =& {|A_+A_-| \over |\Delta|^2} \sin(N\theta +\alpha'),  \\
S_z =& {1 \over 2} {|A_+|^2-|A_-|^2 \over |\Delta|^2}, \nn
\end{align}
where the relative phase $\alpha'=\alpha_--\alpha_+$.
The pseudo-spin vector $(S_x, S_y)$ rotates $N$ times around a vortex with the topological charge $N$ \cite{n.kobayashi}. 

\begin{figure}
\begin{center}
\includegraphics[width=8cm]{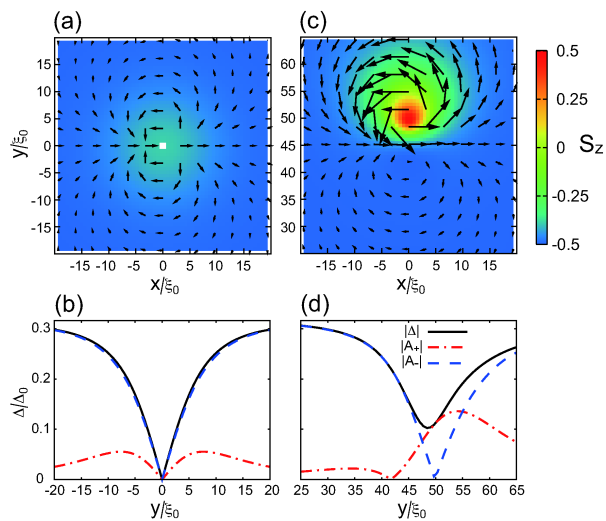}
\end{center}
\caption{(Color online) 
Pseudo-spin vectors $(S_x, S_y)$ with the contour of the $S_z$ component 
(a) around a singular vortex at $\Omega=0.35\Omega_c$ (Fig. \ref{S}) and
(c) around a splitting vortex at $\Omega=0.60\Omega_c$ (Fig. \ref{M}). 
Cross sections of the OP amplitude (solid line), the plus chiral component (dot-dash line), 
and minus chiral component (dashed line) 
(b) across a singular vortex and
(d) across a splitting vortex.
} 
\label{pseudo-spin}
\end{figure}

The axisymmetric singular vortex at the center of the condensate at $\Omega=0.35\Omega_c$ 
has the topological charge $N=2$ (Figs. \ref{pseudo-spin}(a) and \ref{pseudo-spin}(b)).
Similarly, the vortices at $\Omega=0.60\Omega_c$ have the topological charge $N=2$ around a path far away from them 
(Figs. \ref{pseudo-spin}(c) and \ref{pseudo-spin}(d)).
However, they are split into two half-quantum vortices with $N=1$
because of the OP amplitude around the off-centered vortices without axisymmetry by a harmonic trap potential.
The half-quantum vortices on the $y$-axis 
with the combinations of the winding number $\langle -1, 0 \rangle$ and $\langle 0, 1 \rangle$
are located on $(0, 42\xi_0)$ and $(0, 50\xi_0)$, respectively, in Fig. \ref{pseudo-spin}(d).
The distance between two half-quantum vortices is approximately $8\xi_0 \approx 3\xi(t)$,
where $\xi(t)=\xi_0(1-t)^{-1/2}$ is the temperature-dependent coherence length.

Since the off-centered vortex has a finite OP everywhere,
Majorana zero modes, which are zero energy excitations, may not exist.
Note, however, that the vortex situated at a few times of coherence length 
is similar to the singular vortex with Majorana zero modes.
The quasi-particle states of the off-centered vortex are unknown at present.
We must solve the Bogoliubov-de Gennes equations in order to understand the quasi-particle states.
This is our future problem.

\subsection{Trapping dependence}

The number of vortices is proportional to the rotation frequency and the system size, 
or the cross section that is perpendicular to the rotational axis.
The radius of the condensate is inversely proportional to the trap frequency
and is increased by the centrifugal energy under rotation.
Since the effect of the centrifugal energy differs among orbital states,
the region of the polar state appears predominantly at the edge of the condensate under high rotation.
Therefore, the number of vortices in the chiral state situated in the central region does not increase with increasing rotation
because the angular momentum is absorbed by half-quantum vortices in the polar state at the surrounding area,
which prevent the vortex from entering the central region.

The vortex does not enter the chiral state up to high rotation in the case of $\omega=8.0\times 10^{-3}\omega_0$
and only the singular vortex enters under high rotation in the case of $\omega=4.0\times 10^{-3}\omega_0$.
Namely, we will have to employ a low trap frequency in an experiment to observe Majorana zero modes.
In the previous studies under a 3D pancake-shape trap potential \cite{tsutsumi2,tsutsumi3},
the major radius of the condensate was approximately $30\xi_0$,
which corresponds to high-trap-frequency cases in the present study;
therefore, the polar core vortex with a combination of winding number 
$\langle w_+,w_0,w_- \rangle=\langle -1,0,1\rangle$,
which corresponds to the singular vortex $\langle w_+,w_- \rangle=\langle -1,1\rangle$ in this paper,
did not enter the condensate,
where $w_0$ is the winding number of the OP $z$-component, which is absent in the present study.

\section{Conclusions}

We have investigated the stability of the singular vortex that possesses Majorana zero modes at the core
in the $p$-wave superfluidity of the atomic Fermi gases by the GL framework.
As a result, in the systems confined in the quasi-2D harmonic trap potential with a low frequency,
the singular vortex accompanied by Majorana zero modes enters the center of the condensates under rotation.
Under high rotation, the multiple vortex state is stabilized.
The off-centered vortices in the multiple vortex state are split into two half-quantum vortices
and are not accompanied by Majorana zero modes;
the quasi-particle states of the off-centered vortex are unknown at present.
Note that a centered singular vortex is accompanied by Majorana zero modes
even though some vortices exist in the chiral state.

Within the present quasi-2D harmonic trap potential, there is no stable parameter region in the $(\omega,\Omega)$ plane 
for plural singular vortices, which is necessary for a topological quantum computer.
We suggest the use of a quasi-2D square well potential for confinement,
which is also used in the superfluid $^3$He confined in parallel plates \cite{yamashita,kawakami}.
This could accommodate a pair of singular vortices or more.

\section*{Acknowledgements}
We wish to thank M. Ichioka, T. Mizushima, and T. Kawakami for useful discussions.

\end{document}